\documentclass{SciPost}

\hypersetup{
    colorlinks,
    linkcolor={red!50!black},
    citecolor={blue!50!black},
    urlcolor={blue!80!black}
}

\usepackage[bitstream-charter]{mathdesign}
\usepackage{mathtools}
\usepackage{tikz}

\DeclareSymbolFont{usualmathcal}{OMS}{cmsy}{m}{n}
\DeclareSymbolFontAlphabet{\mathcal}{usualmathcal}

\fancypagestyle{SPstyle}{
\fancyhf{}
\lhead{\colorbox{scipostblue}{\bf \color{white} ~SciPost Physics }}
\rhead{{\bf \color{scipostdeepblue} ~Submission }}

\fancyfoot[C]{\textbf{\thepage}}
}

\begin{document}

\pagestyle{SPstyle}

\begin{center}{\Large \textbf{\color{scipostdeepblue}{
Hierarchy of degenerate stationary states in a boundary-driven dipole-conserving spin chain\\
}}}\end{center}

\begin{center}\textbf{
Apoorv Srivastava\textsuperscript{1$\star$} and
Shovan Dutta\textsuperscript{2$\dagger$}
}\end{center}

\begin{center}
{\bf 1} Department of Physics, Indian Institute of Space Science and Technology, Thiruvananthapuram, Kerela, India 695547
\\
{\bf 2} Raman Research Institute, Bangalore 560080, India
\\[\baselineskip]
$\star$ \href{mailto:apoorvsri1909@gmail.com}{\small apoorvsri1909@gmail.com}\,,\quad
$\dagger$ \href{mailto:shovan.dutta@rri.res.in}{\small shovan.dutta@rri.res.in}
\end{center}

\section*{\color{scipostdeepblue}{Abstract}}
\textbf{\boldmath{%
Kinetically constrained spin chains serve as a prototype for structured ergodicity breaking in isolated quantum systems. We show that such a system exhibits a hierarchy of degenerate steady states when driven by incoherent pump and loss at the boundary. By tuning the relative pump and loss and how local the constraints are, one can stabilize mixed steady states, noiseless subsystems, and various decoherence-free subspaces, all of which preserve large amounts of information. We also find that a dipole-conserving bulk suppresses current in steady state. These exact results based on the flow in Hilbert space hold regardless of the specific Hamiltonian or drive mechanism. Our findings show that a competition of kinetic constraints and local drives can induce different forms of ergodicity breaking in open systems, which should be accessible in quantum simulators.
}}

\vspace{\baselineskip}

\noindent\textcolor{white!90!black}{%
\fbox{\parbox{0.975\linewidth}{%
\textcolor{white!40!black}{\begin{tabular}{lr}%
  \begin{minipage}{0.6\textwidth}%
    {\small Copyright attribution to authors. \newline
    This work is a submission to SciPost Physics. \newline
    License information to appear upon publication. \newline
    Publication information to appear upon publication.}
  \end{minipage} & \begin{minipage}{0.4\textwidth}
    {\small Received Date \newline Accepted Date \newline Published Date}%
  \end{minipage}
\end{tabular}}
}}
}


\vspace{10pt}
\noindent\rule{\textwidth}{1pt}
\tableofcontents
\noindent\rule{\textwidth}{1pt}
\vspace{10pt}

\section{Introduction}
\label{sec:intro}

In recent years, kinetic constraint has emerged as an important mechanism to induce layers of ergodicity breaking and subdiffusive transport in isolated many-body quantum systems \cite{Papic_2022_review, Moudgalya_2022_review}. In particular, such constraints can fragment conventional symmetry sectors into exponentially many disjoint blocks, which themselves can vary from ergodic to integrable in the same physical system \cite{Moudgalya_2021_pairhop, Yang_2020_dwall, Hahn_2021_ladder, Rakovszky_2020_sliom, Moudgalya_2022_algebra}. Depending on how local the constraints are, one obtains different levels of fragmentation \cite{Sala_2020_prx, Khemani_2020_prb} and subdiffusion \cite{Feldmeier_2020_subdiff, Morningstar_2020_subdiff, Gromov_2020_hydro, Iaconis_2021_subdiff, Gliozzi_2023_diffhier, Morningstar_2023_longrange}, whose signatures have been observed experimentally \cite{Guo_2021_longrange, Scherg_2021_aidelsburger, Kohlert_2023_aidelsburger, Sanchez_2020_bakr, Adler_2024_2d}. However, the physics is understood only for isolated, at most noisy \cite{Li_2023_open}, systems. Given that one can also engineer local drives in existing setups \cite{Ma_2019_schuster, Li_2025_multiple}, it is particularly timely and intriguing to ask how such a motionally constrained system responds when driven by pump and loss at opposite ends. This is what we address in this paper.

We consider a paradigmatic model for fragmentation, namely a spin chain whose Hamiltonian conserves both a $U(1)$ charge and its dipole moment. These are usually taken as $\smash{\sum_i \hat{S}^z_i}$ and $\smash{\sum_i i \hat{S}^z_i}$, where $i$ labels consecutive sites. The dipole-moment conservation prohibits independent spin exchanges such as $\smash{\hat{S}^+_i \hat{S}^-_j}$, instead admitting pairwise terms such as $\smash{\hat{S}^+_{i-1} \hat{S}^-_{i} \hat{S}^-_{j} \hat{S}^+_{j+1}}$. One can equivalently think of a particle model where the particles hop in pairs to preserve their center of mass. Such correlated hops are realized on strongly tilted lattices \cite{Dai_2017_ringexchange, vanNieuwenburg_2019_pnas, Moudgalya_2021_pairhop, Khemani_2020_prb, Taylor_2020_moessner, Scherg_2021_aidelsburger, Kohlert_2023_aidelsburger, Morningstar_2023_longrange, Guo_2021_longrange, Boesl_2024_bosehubbard} and in quantum Hall setups \cite{Moudgalya_2020_torus, Zerba_2024_fqh}. The allowed separation of the hops ($|i-j|$) controls the fragmentation. For normal spin chains without dipole symmetry, adding boundary drive has proved a fruitful way to explore transport properties and dissipative phase transitions \cite{Landi_2022_boundarydrive}. In the presence of a bias, they typically reach a unique current-carrying steady state. 

In contrast, we show that dipole conservation leads to a degenerate steady-state manifold, which varies qualitatively depending on the level of fragmentation and the nature of the drive. Furthermore, even though the drive breaks dipole conservation locally, the bulk constraints suppress current in steady state, instead giving rise to domain walls. When the Hamiltonian is not fragmented, a fraction of the symmetry sectors are immune to pump and loss at opposite ends and form decoherence-free subspaces (DFSs) \cite{Lidar_1998_DFS, Book_Lidar_Brun} of total size $\sim \exp(\pi \sqrt{L/3})$, where $L$ is the number of spins. With pump and loss at both ends, these flow to one of $L-1$ mixed steady states due to a conserved quantum number or strong symmetry \cite{Buca_Prosen_sym, Baumgartner_2008_sym, Albert_Jiang_sym}. On the other hand, strong fragmentation stabilizes an exponentially large number of DFSs or noiseless subsystems (NSs) \cite{Knill_2000_NS, Book_Lidar_Brun}, where the bulk is shielded from the edges. All of these steady states preserve information \cite{Book_Lidar_Brun, Albert_Jiang_sym} and represent breakdown of ergodicity in the full dynamics \cite{Evans_1977}, which originates from the inability of the Hamiltonian to remove local excitations. Our exact results demonstrate that combining kinetic constraints with local dissipation is a promising route to stabilize different classes of degenerate manifolds in the same setup.

\section{Model and realization}
\label{sec:setup}

For simplicity we consider a spin-$1/2$ chain with nearest-neighbor exchanges, described by the Hamiltonian
\begin{equation}
    \hat{H} = 
    \sum_{i<j} \big( V_{i,j}\; \hat{S}^+_{i-1} \hat{S}^-_{i} \hat{S}^-_{j} \hat{S}^+_{j+1} + \text{h.c.} \big) 
    + \hat{U}\big(\big\{\hat{S}^z_i \big\}\big) \;,
    \label{eq:hamil}
\end{equation}
where $\hat{U}$ is an arbitrary function of $\smash{\hat{S}^z_i}$. We will use the spin and particle languages interchangeably and use $1$ and $0$ to represent spin-$\uparrow$ (occupied site) and spin-$\downarrow$ (empty site), respectively. We focus on two limits: (1) all separations $j-i$ are allowed and there is no fragmentation, i.e., $\hat{H}$ can couple any two Fock states with the same charge and dipole moment, and (2) $V_{i,j} = 0$ unless $j=i+1$, which is strongly fragmented and known as the pair-hopping model \cite{Moudgalya_2021_pairhop}. The dipole moment is conserved locally (in the same sense as used in Ref.~\cite{Gliozzi_2023_diffhier}; see also \cite{Gromov_2020_hydro, Morningstar_2023_longrange}) in (2), which gives subdiffusion, but not in (1), which gives diffusion \cite{Gliozzi_2023_diffhier, Morningstar_2023_longrange}. Our results are based on the connectivity in the Hilbert space and do not rely on the specific forms of $V$ or $\hat{U}$, although they will affect the timescales to reach steady state \cite{Khemani_2020_prb}.

Such models describe the physics of interacting qubits on a strongly tilted lattice \cite{Dai_2017_ringexchange, Moudgalya_2021_pairhop, Khemani_2020_prb, Taylor_2020_moessner, Morningstar_2023_longrange, Guo_2021_longrange} and interactions projected onto the lowest Landau level \cite{Moudgalya_2020_torus, Zerba_2024_fqh}. In particular, the fragmented limit is obtained as the leading-order term for tilted lattices with nearest-neighbor interactions \cite{Moudgalya_2021_pairhop, Khemani_2020_prb, Taylor_2020_moessner} and for a Landau level on a thin torus \cite{Moudgalya_2020_torus}. For long-range interactions or away from the thin-torus limit, one finds other dipole-conserving quartic terms, including longer-range hops \cite{Morningstar_2023_longrange, Guo_2021_longrange}. Nonetheless, here we allow only nearest-neighbor hops, as the DFS structure in the unfragmented case requires finite-range hops. This scenario may be simulated directly using four-qubit interactions among trapped ions \cite{Katz_2023_Nbodyprogrammable, Katz_2023_exp} or digitally with four-qubit gates \cite{Lamata_2018_digital_sc, Fauseweh_2024_digital, Iaconis_2019_automaton, Pai_2019_circuits}.

We assume the spin chain is subjected to incoherent pump and loss at its boundary, which can be modeled most simply by a Lindblad equation \cite{GKS_1976, Lindblad_1976, Breuer_2007_book} for the density matrix $\hat{\rho}$,
\begin{equation}
    \frac{{\rm d}\hat{\rho}}{{\rm d}t} = 
    - {\rm i}\;[\hat{H}, \hat{\rho}] 
    + \sum_{k=1}^4 \left( \hat{L}_k \hat{\rho} \hat{L}_k^{\dagger} - \frac{1}{2} \{\hat{L}_k^{\dagger} \hat{L}_k, \hat{\rho}\} \right) \;,
    \label{eq:lindblad}
\end{equation}
with the jump operators $\hat{L}_1 = \hat{S}^+_1$, $\hat{L}_2 = \hat{S}^-_L$, $\hat{L}_3 = \sqrt{\gamma}\hat{S}^-_1$, and $\hat{L}_4 = \sqrt{\gamma}\hat{S}^+_L$ (we set $\hbar=1$). The dimensionless rate $\gamma$ sets the pump-to-loss ratio. For $\gamma=0$ one has pure pump at one end and pure loss at the other, which we will call ``unipolar'' drive for short.

Such a local Lindblad description is widely used for boundary-driven systems \cite{Landi_2022_boundarydrive} and correspond to coupling the two end sites to infinite-temperature magnetization baths of opposite polarity. It can be physically justified for special types of Markovian baths \cite{Dhahri_2008_lindblad, Landi_2014_lindblad, Schwager_2013_lindblad, Daley_2014_lindblad, Landi_2022_boundarydrive}. However, most of our results are not contingent on this model, or even Markovianity, but only require some form of local injection and removal. Indeed, we use Eq.~\eqref{eq:lindblad} only in Sec.~\ref{sec:strong_sym} for quantifying dynamics in the presence of a strong symmetry. 

While photon loss can be tuned in a lossy cavity \cite{Suleymanzade_2020_tunablecavity}, an incoherent pump was engineered by combining loss and two-photon drive in a superconducting circuit \cite{Ma_2019_schuster}. More generally, one may implement a pump by stochastically measuring $\hat{S}^z_1$ and applying a local $\pi$ pulse only if the outcome was $\downarrow$. In addition, the case of equal pump and loss ($\gamma=1$) corresponds to pure dephasing and can be simulated with noisy $\sigma^x$ and $\sigma^y$ pulses as in Refs.~\cite{Li_2025_multiple, Chenu_2017_noise}.

\section{Unfragmented with unipolar drive: Decoherence-free subspaces}
\label{sec:unfragmented_DFS}

Perhaps the most striking case is when the local nature of the drive and the hops gives rise to multiple DFSs even though the amplitudes $V_{i,j}$ are all nonzero and the model is unfragmented. With pump at the left end and loss at the right end, clearly an empty lattice ends up in $10\dots0$ and a filled lattice in $1\dots10$. In fact, there are $L-1$ frozen configurations with a single domain wall, of the form $1\dots1 0\dots0$, and it is possible to start from a N\'{e}el state such as $1010101$ and end up in $1111000$. Moreover, any superposition of these domain-wall states, which can be strongly entangled, is also frozen. These are not the only final states, however. Configurations of the form $1\dots1010\dots0$ do not evolve either, and pairs of states such as $110010$ and $101100$ oscillate between each other, forming a two-dimensional DFS \cite{Lidar_1998_DFS}. More generally, there are symmetry sectors of $\hat{H}$, with a given particle number $N$ and dipole moment $D$, for which all states have the first site filled and the last site empty. Each of these sectors is unaffected by the pump/loss and becomes a separate DFS. 

To identify these sectors, let us first define $D = \sum_i i n_i$ where $n_i$ are the site occupations. For a given $N$, one can increase $D$ in steps of $1$ by starting from the domain-wall state $1 \dots 1 0 \dots 0$, moving the $N$-th particle to the right until it reaches the right end, then moving the $(N-1)$-th particle to the right, and so on. The last site is empty only during the first sweep, when the state has the form
\begin{equation}
    \underbracket{1 \dots 1}_{N-1} \;
    \underbracket{0 \dots 0}_{p} \; 1 \; 
    \underbracket{0 \dots 0}_{\geq 1} \;\;,
\end{equation}
where $p$ is a nonnegative integer. Starting from such a state, the first site can become $0$ under $\hat{H}$ if the first $N-1$ particles can shift to the right, which requires the last particle to hop $N-1$ sites to the left. Due to the hard-core constraint, this is possible only for $p \geq N$. Hence, the decoherence-free sectors have $p \leq N-1$, corresponding to the ``root'' configurations
\begin{equation}
    \underbracket{1 \dots 1}_{\geq 0} \;
    \underbracket{1\; {\color{red} 1 \dots 1}}_{p} \; 
    \underbracket{{\color{red} 0 \dots 0}}_{p} \;{\color{red} 1} \; 
    \underbracket{0 \dots 0}_{\geq 1}
    \label{eq:DFS_root}
\end{equation}
which reduce to a single domain wall for $p=0$. For $p \geq 2$, only the red-colored sites evolve under the Hamiltonian, forming an active center surrounded by frozen wings. The number of DFSs is given by the number of distinct choices for $(N, p)$, which amounts to $N_{\text{DFS}} = \lfloor L^2/4 \rfloor$, whereas the number of $(N, D)$ sectors is given by $N_{\text{sector}} = (L^3 + 5L + 6)/6$ (see Appendix~\ref{app:unfragmented_DFS}). Thus, only $O(1/L)$ of all the symmetry sectors become DFSs. The size of a DFS is set by the number of arrangements within the active block of the root configuration that have the same dipole moment. This is equivalent to finding integer solutions to the equation
\begin{equation}
    x_1 + x_2 + \dots + x_p = p(p-1)/2 + 2p \;,
    \label{eq:partition_1}
\end{equation}
where $1 \leq x_1 < x_2 < \dots < x_p \leq 2p$. Defining $y_i := x_i - i$ the equation becomes
\begin{equation}
    y_1 + y_2 + \dots + y_p = p \quad \text{with} \quad 0 \leq y_1 \leq y_2 \leq \dots \leq y_p \leq p \;,
    \label{eq:partition_2}
\end{equation}
which are all integer partitions of $p$. Hence, the DFS size grows as $\smash{d_p \sim  \exp(\pi\sqrt{2p/3}) / (4\sqrt{3}p)}$ for $p \gg 1$ \cite{Andrews_1984_paritions}. The total number of all decoherence-free states scales as (see Appendix~\ref{app:unfragmented_DFS})
\begin{equation}
    d_{\text{DFS}} = \sum_{p,N} d_p \sim 
    \frac{\sqrt{3}}{\pi^2} \exp \left(\pi \sqrt{L/3} \right) \left[1 + O(1/\sqrt{L}) \right] \;.
    \label{eq:unfragmented_DFS_size}
\end{equation}

\begin{figure}[!htb]
    \centering
    \includegraphics[width=1\columnwidth]{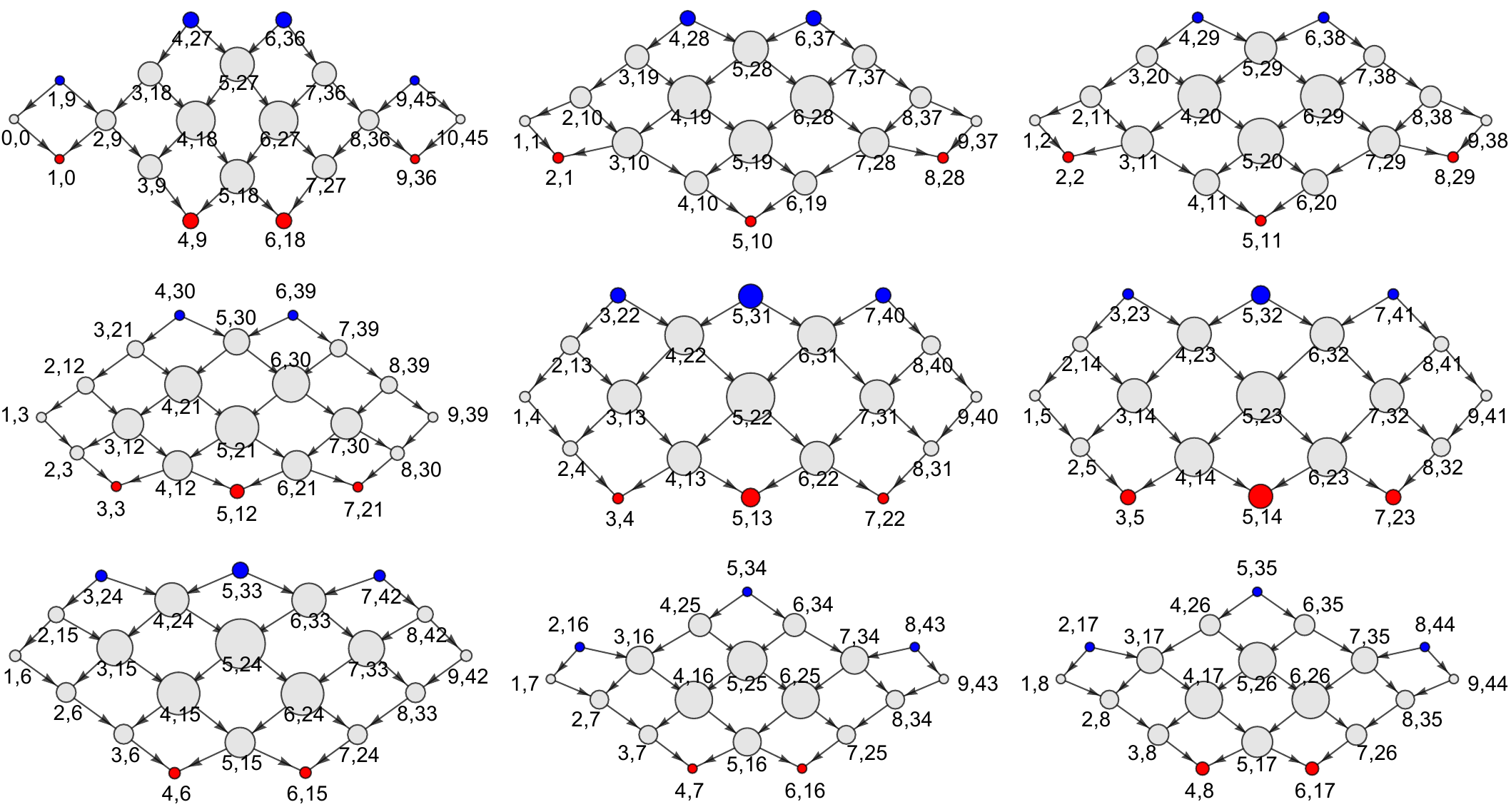}
    \caption{\label{fig:dfs_flow}Flow in the Hilbert space of a dipole-conserving spin-$1/2$ chain with $L=10$ sites, driven by incoherent pump at the first site, $i=0$, and loss at the last site, $i=9$. Each node represents a symmetry sector of the Hamiltonian labeled by the charge $N = \sum_i n_i$ and dipole moment $D = \sum_i i n_i$, where $n_i \in \{0,1\}$ are the site occupations. The area of a node is proportional to its dimension, and the arrows represent either a pump or a loss event. The flow is split into $L-1$ blocks characterized by $D_{\text{mod}} := D \pmod{L-1}$, which varies from $0$ to $8$ from the top left to the bottom right. Inside a block, the flow starts from one of the ``sources'' (blue nodes) at the top and ends in one of the decoherence-free subspaces (red nodes) at the bottom. 
    }
\end{figure}

Not all initial states can reach a given DFS. This is because the dynamics still preserve a quantum number, even though the conservation of $N$ and $D$ are broken. If we call the leftmost site $i=0$, the pump does not alter $D$ and the loss at the right end decreases $D$ only in steps of $L-1$. Thus, they conserve the integer $D_{\text{mod}} := D \pmod{L-1}$, splitting the dynamics into $L-1$ blocks of roughly equal size. The resulting flow in Hilbert space in shown for $L=10$ in Fig.~\ref{fig:dfs_flow}, where each node represents a symmetry sector of the Hamiltonian, labeled by $(N,D)$, and each arrow represents a pump or loss event connecting two sectors. All paths flow to one of the DFSs, which are colored red. Conversely, the blue-colored ``sources'' have the first site empty and the last site filled---these would become DFSs if one were to swap the pump and loss. As exchanging left and right changes $D$ to $(L-1)N - D$, the sources for a given $D_{\text{mod}}$ turn into the sinks for $L - 1 - D_{\text{mod}}$. For $D_{\text{mod}} = 0$ the two are balanced.

We make two observations: First, the arrows show classical paths as they denote either a pump or a loss event. Therefore, starting with a definite $(N,D)$ one always ends up in a unique DFS. In order to reach a superposition of multiple DFSs, such as two of the domain-wall states, one must necessarily start from a superposition of multiple nodes. So the dissipative drive does not provide a straightforward way to produce structured entanglement. Of course, a product state can still end up in a large DFS with volume-law entanglement \cite{Moudgalya_2022_review, Moudgalya_2021_pairhop, Yang_2020_dwall, Hahn_2021_ladder, Moudgalya_2021_spectrum, Herviou_2021_mbl, Sala_2020_prx, Morningstar_2020_subdiff, Pozderac_2023_ergidicity, Classen_2024_thermalization}. Second, in all of the DFSs the first and last sites are frozen at $1$ and $0$, respectively. Thus, there is no net current in steady state and any unidirectional flow must be transient.

\section{Unfragmented with bipolar drive: Strong symmetry}
\label{sec:strong_sym}

\begin{figure}[!h]
    \centering
    \includegraphics[width=1\columnwidth]{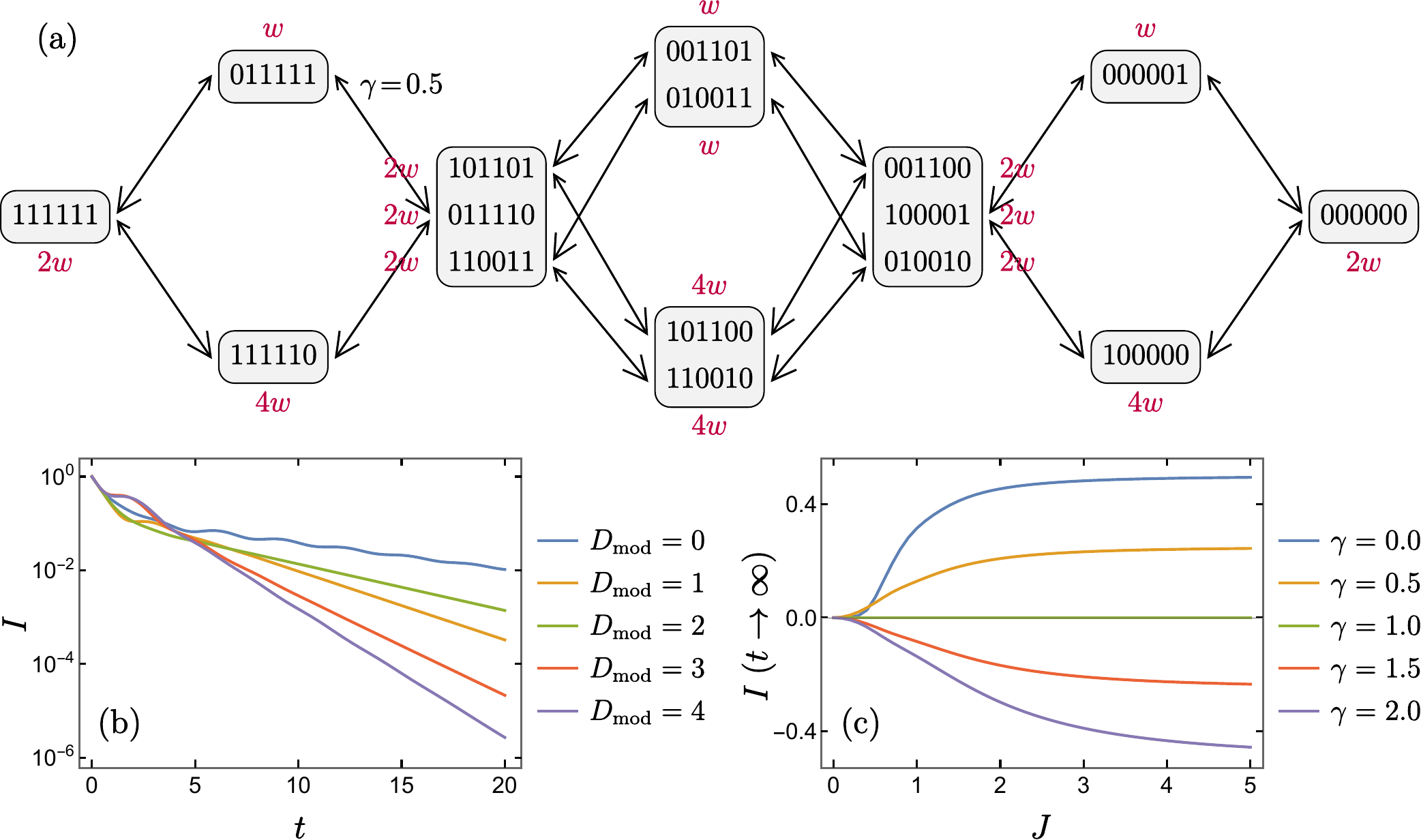}
    \caption{\label{fig:strong_sym}(a) Flow in the $D_{\text{mod}} = 0$ block for $L=6$ in the presence of pump and loss at both ends. Downward transitions are caused by injection at the left end or loss at the right end (with rate $1$), whereas upward transitions are caused by loss at the left end or injection at the right end with rate $\gamma = 0.5$. The red labels show steady-state population of each configuration (up to normalization), satisfying detailed balance. (b) Overall rate at which particles are injected into the first site as a function of time for the Lindblad dynamics in Eq.~\eqref{eq:lindblad} with $L=6$, $\gamma=0.5$, $V_{i,j} = 1$, and $\hat{U} = 0$, starting from a single particle at site $i = D_{\text{mod}}$. (c) The same rate in steady state after adding the symmetry-breaking term $\delta\hat{H} = -J \sum_i \hat{S}^+_i \hat{S}^-_{i+1} + \text{h.c.}$ to the Hamiltonian.
    }
\end{figure}

With pump and loss at both ends, the arrows in Fig.~\ref{fig:dfs_flow} become bidirectional. Hence, the DFSs are no longer stable. However, the conservation of $D_{\text{mod}}$ still holds. Such a quantum number, which is conserved by both the Hamiltonian and the dissipation, is called a strong symmetry in the context of Lindblad dynamics \cite{Buca_Prosen_sym, Baumgartner_2008_sym, Albert_Jiang_sym}. In the absence of further symmetries, each block of $D_{\text{mod}}$ reaches a unique, mixed steady state. The structure of the steady state also follows from the flow in Fock space. In particular, we see that the Fock states can be arranged vertically in layers such that the pump and loss couple only adjacent layers. If, in addition, the pump rate at the first site equals the loss rate at the last site (and vice versa) \cite{Huber_2020_PT, Prosen_2012_PT, ElGanainy_2018_PT}, all transitions from a given layer to the one below it have the same rate (set to 1), and all upward transitions have rate $\gamma$ [see Eq.~\eqref{eq:lindblad}]. Such a flow diagram implies a steady state with detailed balance \cite{Agarwal_1973_detbal}, where all configurations in a given layer are equally likely and the weights in successive layers differ by a factor of $\gamma$, as sketched in Fig.~\ref{fig:strong_sym}(a) for $L=6$ and $\gamma=0.5$. Thus, each $(N,D)$ sector heats to infinite temperature, and their relative weights ensure no net flow between any two Fock states. To further characterize these weights, note that the successive nodes in a given layer differ in $(N,D)$ by $(2, L-1)$. Thus, all of them share the same integer
\begin{equation}
    Q := N - 2\; (D - D_{\text{mod}})/(L-1) \;,
    \label{eq:horizontal_level}
\end{equation}
which increases in steps of $1$ from the top layer to the bottom layer. The steady state is given by the Gibbs ensemble $\hat{\rho} \propto \exp(\mu \hat{Q})$ with $\mu = \ln(1/\gamma)$. This holds irrespective of the form of the Hamiltonian as long as it couples all Fock states with the same $N$ and $D$.

As one might expect, there is no net current in the system in such an infinite-temperature steady state. This is confirmed explicitly for the Lindblad dynamics [Eq.~\eqref{eq:lindblad}] for small systems. As shown in Fig.~\ref{fig:strong_sym}(b) for $V_{i,j}=1$, the overall rate of injection from outside into the first site, $I = 1 - \langle \hat{n}_0 \rangle - \gamma \langle \hat{n}_0 \rangle$, decays exponentially with time for all values of $D_{\text{mod}}$. The decay rate is set by the Liouvillian gap which would depend on $V_{i,j}$, $\gamma$, and $L$ \cite{Znidaric_2015_gap, Can_2019_gap}. We also find that the current is restored by breaking the dipole symmetry of the Hamiltonian, e.g., with a magnetic field in the $x$ direction or by independent nearest-neighbor hops. Figure~\ref{fig:strong_sym}(c) shows how the steady-state current grows monotonically with the strength of the latter.

\section{Fragmented with unipolar drive: Decoherence-free subspaces}
\label{sec:fragmented_DFS}

Next we turn to the pair-hopping model \cite{Moudgalya_2021_pairhop}, for which $V_{i,j} = 0$ unless $j=i+1$ and each $(N,D)$ sector is strongly fragmented. Due to the nearest-neighbor constraint, any state composed of sequences of three or more $1$'s or $0$'s does not evolve under the Hamiltonian. Thus, with pump at left end and loss at right end, we obtain exponentially many frozen states of the form
\begin{equation}
    \underbracket{1 \dots 1}_{\geq 3} \; 
    \underbracket{0 \dots 0}_{\geq 3} \;
    \underbracket{1 \dots 1}_{\geq 3} \dots 
    \underbracket{0 \dots 0}_{\geq 3} \;.
\end{equation}
This is in stark contrast with the unfragmented case, where only $2L-4$ configurations, with one or three domain walls, were frozen. Furthermore, as we show below, there is a much larger set of decoherence-free states, including $2^{L-4}$ configurations of the form 
\begin{equation}
    11\;\fbox{\phantom{Arbitrary sequence}}\;00 \;,
    \label{eq:fragmented_bulk_DFS}
\end{equation}
where the box can contain an arbitrary binary sequence. These states are grouped into exponentially many DFSs given by the fragmentation structure of the Hamiltonian \cite{Moudgalya_2021_pairhop}.

The form in Eq.~\eqref{eq:fragmented_bulk_DFS} is not necessary for a state to be decoherence free. All we need is that the first site is frozen at $1$ and the last site is frozen at $0$. Thus, it suffices to discard only those configurations of the form $1\;\fbox{\phantom{Arbitrary}}\;0$ for which this is not true. To see how this works, note that the dynamics under the Hamiltonian consist of changing four consecutive sites from $1001$ to $0110$ and vice versa. Hence, the first spin can flip only if the first four sites are $1001$, which means we should omit states of the form $1001\;\fbox{\phantom{Arbitrary}}\;0$. However, these can in turn arise from $101001\;\fbox{\phantom{Arbitrary}}\;0$ and $1000110\;\fbox{\phantom{Arbitrary}}\;0$, and so on. This hierarchy of decohering states is generated by recursively applying the expansion rules
\begin{equation}
    \begin{minipage}{0.8\textwidth}
    \centering
    \begin{tikzpicture}
        \node at (-2,0) (label1) {$\underline{01}$};
        \node at (-3,-1) (nodeA1) {$10\underline{01}$};
        \node at (-1,-1) (nodeB1) {$001\underline{10}$};
        
        \draw[->, line width=1pt] (label1) -- (nodeA1);
        \draw[->, line width=1pt] (label1) -- (nodeB1);
        
        \node at (2,0) (label2) {$\underline{10}$};
        \node at (1,-1) (nodeA2) {$01\underline{10}$};
        \node at (3,-1) (nodeB2) {$110\underline{01}$};

        \draw[->, line width=1pt] (label2) -- (nodeA2);
        \draw[->, line width=1pt] (label2) -- (nodeB2);
    \end{tikzpicture}
    \end{minipage}
    \label{eq:expand_right}
\end{equation}
to the last two digits (underlined) of the lead sequence. The resulting ``1001 family'' of states have the form
\begin{subequations}
\begin{align}
    \Uparrow \dots \Uparrow 
    0 \Downarrow \dots \Downarrow 
    1 \Uparrow \dots \Uparrow 
    0 \Downarrow \dots \Downarrow 
    1 \cdots \; \cdots 
    1 \Uparrow \dots {\color{red}\Uparrow \Downarrow} \; 
    \fbox{\phantom{Arbitrary}} \; 0 & 
    \label{eq:1001_family_1} \\[2pt]
    \text{or} \quad 
    \Uparrow \dots \Uparrow 
    0 \Downarrow \dots \Downarrow 
    1 \Uparrow \dots \Uparrow 
    0 \Downarrow \dots \Downarrow 
    1 \cdots \; \cdots 
    0 \Downarrow \dots {\color{red}\Downarrow \Uparrow} \; 
    \fbox{\phantom{Arbitrary}} \; 0 & \;, 
    \label{eq:1001_family_2}
\end{align}
\label{eq:1001_family}
\end{subequations}
where the collective spins $\Uparrow$ and $\Downarrow$ stand for $10$ and $01$, respectively \cite{Moudgalya_2021_pairhop}. The last two collective spins (colored red) represent the active edge of the sequence---no site to their left can change without first swapping them. Such a swap produces an ancestor in the family tree. As we detail in Appendix~\ref{app:fragmented_DFS}, the total number of states in this family, for which the first site can eventually flip, is $N_{1001} \approx 2^{L-2}/5$. Similarly, the last site can flip if the last four sites become $0110$, which in turn gives rise to a family of states expanding to the left of the same size as $N_{1001}$. The set of all decoherence-free states is obtained by discarding members of either of the two families, which leaves a manifold of total size (see Appendix~\ref{app:fragmented_DFS})
\begin{equation}
    d_{\text{DFS}} \approx (2/5)^2 \times 2^L \;.
    \label{eq:fragmented_DFS_size}
\end{equation}

In summary, the bulk of the system remains strongly fragmented and a finite fraction of the Hilbert space becomes decoherence free, as opposed to a vanishingly small fraction in the unfragmented case [Eq.~\eqref{eq:unfragmented_DFS_size}]. As in the latter, there is no net current in steady state.

\section{Fragmented with bipolar drive: Noiseless subsystems}
\label{sec:fragmented_NS}

With pump and loss at both ends, no state is unaffected by the drive, so we do not have a DFS. However, we can insert a ``blockade'' sequence \cite{Moudgalya_2021_pairhop, Khemani_2020_prb, Sala_2020_prx, Classen_2024_thermalization} of $1111$ or $0000$ at either end to shield the bulk from the boundary. These are initial states of the form
\begin{equation}
    \fbox{0/1} \; 111 \; 
    \fbox{1\phantom{Arbitrary sequence}0} \;
    000 \; \fbox{0/1} \;.
    \label{eq:blockade_NS}
\end{equation}
From our discussion in Sec.~\ref{sec:fragmented_DFS} and the nearest-neighbor constraint, it follows that the first four and last four sites cannot change under $\hat{H}$. Thus, information in the bulk is preserved, which constitutes a noiseless subsystem (NS) \cite{Book_Lidar_Brun, Knill_2000_NS}, while the end sites continue to flip between $0$ and $1$, taking the system to a mixed steady state [see Eq.~\eqref{eq:lindblad}]
\begin{equation}
    \hat{\rho} \propto 
    \big( \gamma |0\rangle\langle0| + |1\rangle\langle1| \big) 
    \otimes 
    \big|\psi^{\text{bulk}}_j \big\rangle 
    \big\langle \psi^{\text{bulk}}_j \big| 
    \otimes 
    \big( |0\rangle\langle0| + \gamma |1\rangle\langle1| \big) \;,
\end{equation}
where $\smash{|\psi^{\text{bulk}}_j\rangle}$ is an eigenstate of the bulk Hamiltonian. In fact, the number of bulk states that are immune to the end flips is much larger than the $2^{L-6}$ configurations of the type in Eq.~\eqref{eq:blockade_NS}. All we require is that the two end sites are not altered by $\hat{H}$. As we saw in Sec.~\ref{sec:fragmented_DFS}, the first site is not altered from $1$ for all states outside the $1001$ family [Eq.~\eqref{eq:1001_family}], and the last site is not altered from $0$ for all states outside the $0110$ family [Eq.~\eqref{eq:0110_family}]. One has to also exclude states for which the Hamiltonian can flip the first site from $0$ to $1$ or the last site from $1$ to $0$. These are obtained by swapping $1$'s with $0$'s in the two previous families. By symmetry the four groups are of equal size. They comprise a total of approximately $(16/25) \; 2^{L-2}$ configurations, as sketched below (with arrows indicating the direction in which the family expands).
\begin{equation}
    \begin{minipage}{0.8\textwidth}
    \centering
    \usetikzlibrary{patterns}
    \begin{tikzpicture}
        \draw[line width=1pt] (-0.8,0.8) circle (1) node[shift={(-6pt,6pt)}, anchor=center, align=center] {$A:$ \\ $1001\!\rightarrow$};
        \draw[line width=1pt] (0.8,0.8) circle (1) node[shift={(6pt,6pt)}, anchor=center, align=center] {$B:$ \\ $\leftarrow\!0110$};
        \draw[line width=1pt] (0.8,-0.8) circle (1) node[shift={(6pt,-3pt)}, anchor=center, align=center] {$C:$ \\ $0110\!\rightarrow$};
        \draw[line width=1pt] (-0.8,-0.8) circle (1) node[shift={(-6pt,-3pt)}, anchor=center, align=center] {$D:$ \\ $\leftarrow\!1001$};
        \begin{scope}
            \clip (-0.8,0.8) circle (1);
            \fill[pattern=north east lines] (0.8,0.8) circle (1);
        \end{scope}
        \begin{scope}
            \clip (-0.8,-0.8) circle (1);
            \fill[pattern=north east lines] (0.8,-0.8) circle (1);
        \end{scope}
        \begin{scope}
            \clip (-0.8,-0.8) circle (1);
            \fill[pattern=north east lines] (-0.8,0.8) circle (1);
        \end{scope}
        \begin{scope}
            \clip (0.8,-0.8) circle (1);
            \fill[pattern=north east lines] (0.8,0.8) circle (1);
        \end{scope}
        \node[anchor=west] at (2.5, 0.4) {$d_A = d_B = d_C = d_D \approx 2^{L-2}/5 \;,$};
        \node[anchor=west] at (2.5, -0.4) {$d_{A \cap B} = d_{C \cap D} \approx d_{A \cap D} = d_{B \cap C} \approx 2^{L-2}/25 \;.$};
    \end{tikzpicture}
    \end{minipage}
\end{equation}
Hence, the dimension of the NS is given by $d_{\text{NS}} \approx (9/25) \; 2^{L-2}$. These bulk states follow the fragmentation of the Hamiltonian. As before, there is no net current in steady state.

\section{Conclusion}
\label{sec:conclusion}

We studied how a dipole-conserving spin-$1/2$ chain responds when driven by incoherent pump and loss at opposite ends. We found that, even though the drive locally breaks both charge and dipole conservation, the kinetic constraints in the bulk suppress current at late times, driving the system to a degenerate steady-state manifold. By analyzing the flow in Hilbert space, we were able to derive exact results that fully characterize the manifold and does not depend on the precise form of the Hamiltonian or the details of how the pump and loss are implemented. We showed the nature of the manifold can be tuned across a wide range of possibilities (DFS, NS, mixed steady states) by varying whether the dipole conservation is local or global and whether the pump and loss act on opposite ends or both ends. Such flexibility is highly unusual for Markovian systems which generically evolve to a unique steady state \cite{Evans_1977, Albert_Jiang_sym}. Our main findings should be accessible in present-day quantum simulators (see Sec.~\ref{sec:setup}). Our results show that combining kinetic constraints with local dissipative drives is a promising avenue to induce different types of ergodicity breaking in open quantum dynamics.

We conclude with three remarks for future work. First, while we have focused on steady states, past studies have established an intimate link between multipole conservation and subdiffusion in isolated systems \cite{Feldmeier_2020_subdiff, Morningstar_2020_subdiff, Gromov_2020_hydro, Iaconis_2021_subdiff, Gliozzi_2023_diffhier, Morningstar_2023_longrange}, which begs the question of how the subdiffusive transport controls the approach to steady state \cite{Landi_2022_boundarydrive}. Second, although we have taken a boundary-driven spin-$1/2$ chain for simplicity, we expect our conclusions to generalize to other cases. Nonetheless, it would be particularly interesting to see whether one can stabilize a current-carrying steady state by increasing the local dimension \cite{Sala_2020_prx, Khemani_2020_prb, Morningstar_2023_longrange, Scherg_2021_aidelsburger, Boesl_2024_bosehubbard, Classen_2024_thermalization} or by driving interior sites \cite{Dutta_2021_interior}. Third, as dipole-conserving models are fragmented in a product basis \cite{Sala_2020_prx, Khemani_2020_prb, Moudgalya_2022_algebra}, most of our findings carry over to a dipole-conserving classical exclusion process \cite{Ritort_2003_KCM, Cancrini_2007_KCSM, Garrahan_2011_KCM, Feldmeier_2020_subdiff, Iaconis_2019_automaton, Pai_2019_circuits, Gopalakrishnan_2018_automata, Mukherjee_2024_hyperuniformity, Hazra_2025_hyperuniformity}. But, for the same reason, it is difficult to create entangled states by a local drive (see Sec.~\ref{sec:unfragmented_DFS}). It would be valuable to explore if one can circumvent this drawback in systems that are fragmented in an entangled basis \cite{Moudgalya_2022_algebra, Li_2023_open} or by using fermionic loss \cite{Dutta_2020_center}.

\section*{Acknowledgements}
We thank Sanjay Moudgalya for useful discussions.

\paragraph{Funding information}
This work was supported by intramural funds of the Raman Research Institute.

\begin{appendix}
\numberwithin{equation}{section}

\section{Counting decoherence-free states in the unfragmented case}
\label{app:unfragmented_DFS}

For a given particle number $N$, the dipole moment $D$ is minimum for the state $1 \dots 1 0 \dots 0$ and increases in steps of $1$ as we move each particle $L-N$ sites to the right, one at a time. Hence, the number of distinct $(N, D)$ sectors is given by $N_{\text{sector}} = \sum_{N=0}^L 1 + N(L-N) = (L^3 + 5L + 6)/6$.

The number of disjoint DFSs can be found by counting configurations of the type in Eq.~\eqref{eq:DFS_root} with $p \leq N-1$ that can fit in $L$ sites, i.e., $N+p+1 \leq L$. Thus,
\begin{equation}
    N_{\text{DFS}} 
    = \sum_{N=1}^{L-1} \sum_{p = 0}^{\text{min}(N,L-N)-1} 1 
    = \left\lfloor \frac{L^2}{4} \right\rfloor \;.
\end{equation}

As shown in Eqs.~\eqref{eq:partition_1} and \eqref{eq:partition_2}, the size of a DFS is given by the number of integer partitions of $p$, denoted by $d_p$. Hence, the total count of all decoherence-free states amounts to
\begin{equation}
    d_{\text{DFS}} 
    = \sum_{N=1}^{L-1} \sum_{p = 0}^{\text{min}(N,L-N)-1} d_p 
    = \sum_{p=0}^{\lfloor L/2 \rfloor -1} \sum_{N=p+1}^{L-p-1} d_p
    = \sum_{p=0}^{\lfloor L/2 \rfloor -1} (L-2p-1) d_p \;.
\end{equation}
Using $d_p \sim \exp(\pi\sqrt{2p/3}) / (4\sqrt{3}p)$ for $p \gg 1$ \cite{Andrews_1984_paritions}, we find
\begin{equation}
    d_{\text{DFS}} \sim 
    \int_1^{L/2} {\rm d}p \;(L-2p) \frac{e^{\pi\sqrt{2p/3}}}{4\sqrt{3}p} 
    = F(L) - F(2) \;,
\end{equation}
where 
\begin{equation}
    F(x) \coloneqq 
    \frac{L}{2\sqrt{3}} \text{ Ei} \bigg(\pi \sqrt{\frac{x}{3}} \bigg) 
    - e^{\pi \sqrt{x/3}} \bigg( \frac{\sqrt{x}}{2\pi} - \frac{\sqrt{3}}{2\pi^2} \bigg) \;
\end{equation}
and Ei is the exponential integral function \cite{Book_NIST}. Using the asymptotic expansion
\begin{equation}
    \text{Ei}(x) \sim \frac{e^x}{x} \left(1 + \frac{1!}{x} + \frac{2!}{x^2} + \cdots \right)
\end{equation}
gives the scaling
\begin{equation}
    d_{\text{DFS}} \sim \frac{\sqrt{3}}{\pi^2} e^{\pi \sqrt{L/3}} \left[ 1 + \frac{\sqrt{3}}{\pi \sqrt{L}} + O\left(\frac{1}{L}\right) \right] \;.
\end{equation}

\section{Counting decoherence-free states in the fragmented case}
\label{app:fragmented_DFS}

First we count the $1001$ family given by the configurations in Eq.~\eqref{eq:1001_family}. The lead sequence of such a state can be labeled the number of collective spins, $n \geq 2$, and the number of domain walls of the form $\Uparrow 0 \Downarrow$ and $\Downarrow 1 \Uparrow$, $k \leq n-2$. For a finite system they also satisfy $2n+k \leq L-1$. There are $\binom{n-2}{k}$ ways of placing the domain walls and $2^{L-1-2n-k}$ arrangements of the sites in the box for a given lead sequence [see Eq.~\eqref{eq:1001_family}]. Hence, the total number of states is given by
\begin{equation}
    N_{1001} = 
    \sum_{n=2}^{\lfloor (L-1)/2 \rfloor} \;
    \sum_{k=0}^{\min(n-2, L-1-2n)} 
    \binom{n-2}{k} \; 2^{L-1-2n-k}
    \; \approx \frac{2^{L-2}}{5} \;.
    \label{eq:N_1001}
\end{equation}
For odd values of $L \geq 7$, one has to add states for which the lead sequence spans the lattice, ending with $0110$. These have an odd number of domain walls that can placed in $(L-k)/2-2$ positions, yielding a total count of approximately $0.13 \times 1.32^L$.

\begin{figure}[!b]
    \centering
    \includegraphics[width=1\columnwidth]{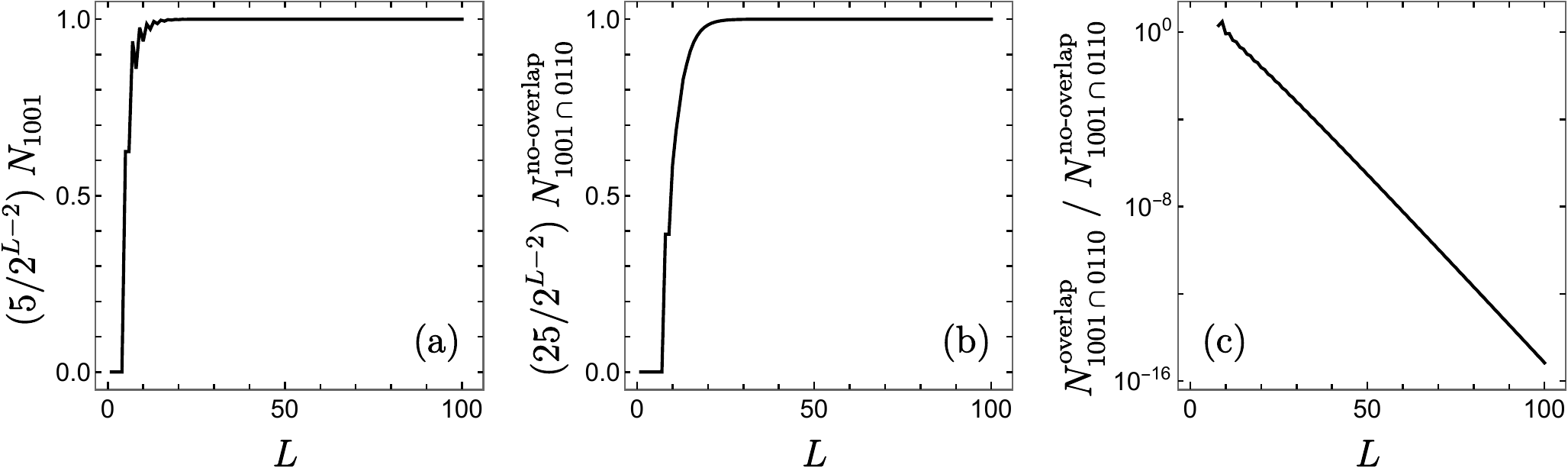}
    \caption{\label{fig:dfs_count}(a) Number of states in the $1001$ family, having configurations of the form in Eq.~\eqref{eq:1001_family} for which the first site can flip from $1$ to $0$ through a sequence of swaps between $1001$ to $0110$. (b) Number of states of the form in Eq.~\eqref{eq:nonoverlapping_states}, for which both the first site and the last site can flip from $1$ and $0$, respectively. (c) Number of such unstable states for which the lead sequences on either side overlap, in comparison. 
    }
\end{figure}

Next we turn to the $0110$ family, for which the last site can eventually flip from $0$ to $1$. Like the $1001$ family, the states are generated by recursively applying the expansion rules
\begin{equation}
    \begin{minipage}{0.8\textwidth}
    \centering
    \begin{tikzpicture}
        \node at (-2,0) (label1) {$\underline{01}$};
        \node at (-3,-1) (nodeA1) {$\underline{01}10$};
        \node at (-1,-1) (nodeB1) {$\underline{10}011$};
        
        \draw[->, line width=1pt] (label1) -- (nodeA1);
        \draw[->, line width=1pt] (label1) -- (nodeB1);
        
        \node at (2,0) (label2) {$\underline{10}$};
        \node at (1,-1) (nodeA2) {$\underline{10}01$};
        \node at (3,-1) (nodeB2) {$\underline{01}100$};

        \draw[->, line width=1pt] (label2) -- (nodeA2);
        \draw[->, line width=1pt] (label2) -- (nodeB2);
    \end{tikzpicture}
    \end{minipage}
    \label{eq:expand_left}
\end{equation}
to the leftmost two digits (underlined), which leads to the form
\begin{subequations}
\begin{align}
    1 \; \fbox{\phantom{Arbitrary}} \;
    {\color{red}\Uparrow \Downarrow} \dots \Downarrow 1 
    \cdots \; \cdots 
    0 \Downarrow \dots \Downarrow 
    1 \Uparrow \dots \Uparrow 
    0 \Downarrow \dots \Downarrow 
    1 \Uparrow \dots \Uparrow & 
    \label{eq:0110_family_1} \\[2pt]
    \text{or} \quad 
    1 \; \fbox{\phantom{Arbitrary}} \;
    {\color{red}\Downarrow \Uparrow} \dots \Uparrow 0 
    \cdots \; \cdots 
    0 \Downarrow \dots \Downarrow 
    1 \Uparrow \dots \Uparrow 
    0 \Downarrow \dots \Downarrow 
    1 \Uparrow \dots \Uparrow & \;, 
    \label{eq:0110_family_2}
\end{align}
\label{eq:0110_family}
\end{subequations}
By symmetry the two families have equal size, $N_{0110} = N_{1001} \approx 2^{L-2}/5$ [see Fig.~\ref{fig:dfs_count}(a)].

The number of states shared between the two is dominated by configurations of the form
\begin{equation}
    \Uparrow \dots \Uparrow 
    0 \Downarrow \dots \Downarrow 
    1 \cdots \; \cdots 
    {\color{red}\Uparrow \Downarrow} \; 
    \fbox{\phantom{Arbitrary sequence}} \;
    {\color{red}\Downarrow \Uparrow} 
    \cdots \; \cdots 
    0 \Downarrow \dots \Downarrow 
    1 \Uparrow \dots \Uparrow \;,
    \label{eq:nonoverlapping_states}
\end{equation}
where the lead sequences do not overlap. Labeling the number of collective spins in these two sequences by $m$ and $n$, and the number of domain walls by $p$ and $q$, we find [as in Eq.~\eqref{eq:N_1001}]
\begin{equation}
    N_{1001 \cap 0110}^{\text{no-overlap}} = 
    \underset{2(m+n)+p+q \leq L}{
    \sum_{m,n \geq 2} \;
    \sum_{p=0}^{m-2} \;
    \sum_{q=0}^{n-2}
    } \;\;
    \binom{m-2}{p} \binom{n-2}{q} \;
    2^{L-2(m+n)-p-q} \;
    \approx \frac{2^{L-2}}{25} \;.
\end{equation}

For the other shared states, the lead sequences overlap by at most $4$ sites. This is because the patterns $1001$ and $0110$ only appear at the active edges (red collective spins). As we show in Fig.~\ref{fig:dfs_count}(c), this overlapping set is exponentially small compared to the non-overlapping set. Hence, the total size of the two families can be estimated as
\begin{equation}
    N_{1001 \cup 0110} \approx 
    N_{1001} + N_{0110} - N_{1001 \cap 0110}^{\text{no-overlap}} 
    \approx (9/25) \; 2^{L-2} \;.
\end{equation}

\end{appendix}




\begin{thebibliography}{10}
\providecommand{\url}[1]{\texttt{#1}}
\providecommand{\urlprefix}{URL }
\expandafter\ifx\csname urlstyle\endcsname\relax
  \providecommand{\doi}[1]{doi:\discretionary{}{}{}#1}\else
  \providecommand{\doi}{doi:\discretionary{}{}{}\begingroup \urlstyle{rm}\Url}\fi
\providecommand{\eprint}[2][]{\url{#2}}

\bibitem{Papic_2022_review}
Z.~Papić,
\newblock \emph{Weak ergodicity breaking through the lens of quantum entanglement},
\newblock In \emph{Entanglement in Spin Chains}, p. 341–395. Springer, Cham, Switzerland,
\newblock ISBN 9783031039980,
\newblock \doi{10.1007/978-3-031-03998-0_13} (2022).

\bibitem{Moudgalya_2022_review}
S.~Moudgalya, B.~A. Bernevig and N.~Regnault,
\newblock \emph{Quantum many-body scars and {H}ilbert space fragmentation: a review of exact results},
\newblock Rep. Prog. Phys. \textbf{85}, 086501 (2022),
\newblock \doi{10.1088/1361-6633/ac73a0}.

\bibitem{Moudgalya_2021_pairhop}
S.~Moudgalya, A.~Prem, R.~Nandkishore, N.~Regnault and B.~A. Bernevig,
\newblock \emph{Thermalization and its absence within {K}rylov subspaces of a constrained {H}amiltonian},
\newblock In \emph{Memorial Volume for Shoucheng Zhang}, p. 147–209. World Scientific,
\newblock ISBN 9789811231711,
\newblock \doi{10.1142/9789811231711_0009} (2021).

\bibitem{Yang_2020_dwall}
Z.-C. Yang, F.~Liu, A.~V. Gorshkov and T.~Iadecola,
\newblock \emph{Hilbert-space fragmentation from strict confinement},
\newblock Phys. Rev. Lett. \textbf{124}, 207602 (2020),
\newblock \doi{10.1103/physrevlett.124.207602}.

\bibitem{Hahn_2021_ladder}
D.~Hahn, P.~A. McClarty and D.~J. Luitz,
\newblock \emph{Information dynamics in a model with {H}ilbert space fragmentation},
\newblock SciPost Phys. \textbf{11}, 074 (2021),
\newblock \doi{10.21468/scipostphys.11.4.074}.

\bibitem{Rakovszky_2020_sliom}
T.~Rakovszky, P.~Sala, R.~Verresen, M.~Knap and F.~Pollmann,
\newblock \emph{Statistical localization: From strong fragmentation to strong edge modes},
\newblock Phys. Rev. B \textbf{101}, 125126 (2020),
\newblock \doi{10.1103/physrevb.101.125126}.

\bibitem{Moudgalya_2022_algebra}
S.~Moudgalya and O.~I. Motrunich,
\newblock \emph{Hilbert space fragmentation and commutant algebras},
\newblock Phys. Rev. X \textbf{12}, 011050 (2022),
\newblock \doi{10.1103/physrevx.12.011050}.

\bibitem{Sala_2020_prx}
P.~Sala, T.~Rakovszky, R.~Verresen, M.~Knap and F.~Pollmann,
\newblock \emph{Ergodicity breaking arising from {H}ilbert space fragmentation in dipole-conserving {H}amiltonians},
\newblock Phys. Rev. X \textbf{10}, 011047 (2020),
\newblock \doi{10.1103/physrevx.10.011047}.

\bibitem{Khemani_2020_prb}
V.~Khemani, M.~Hermele and R.~Nandkishore,
\newblock \emph{Localization from {H}ilbert space shattering: From theory to physical realizations},
\newblock Phys. Rev. B \textbf{101}, 174204 (2020),
\newblock \doi{10.1103/physrevb.101.174204}.

\bibitem{Feldmeier_2020_subdiff}
J.~Feldmeier, P.~Sala, G.~De~Tomasi, F.~Pollmann and M.~Knap,
\newblock \emph{Anomalous diffusion in dipole- and higher-moment-conserving systems},
\newblock Phys. Rev. Lett. \textbf{125}, 245303 (2020),
\newblock \doi{10.1103/physrevlett.125.245303}.

\bibitem{Morningstar_2020_subdiff}
A.~Morningstar, V.~Khemani and D.~A. Huse,
\newblock \emph{Kinetically constrained freezing transition in a dipole-conserving system},
\newblock Phys. Rev. B \textbf{101}, 214205 (2020),
\newblock \doi{10.1103/physrevb.101.214205}.

\bibitem{Gromov_2020_hydro}
A.~Gromov, A.~Lucas and R.~M. Nandkishore,
\newblock \emph{Fracton hydrodynamics},
\newblock Phys. Rev. Res. \textbf{2}, 033124 (2020),
\newblock \doi{10.1103/physrevresearch.2.033124}.

\bibitem{Iaconis_2021_subdiff}
J.~Iaconis, A.~Lucas and R.~Nandkishore,
\newblock \emph{Multipole conservation laws and subdiffusion in any dimension},
\newblock Phys. Rev. E \textbf{103}, 022142 (2021),
\newblock \doi{10.1103/physreve.103.022142}.

\bibitem{Gliozzi_2023_diffhier}
J.~Gliozzi, J.~May-Mann, T.~L. Hughes and G.~De~Tomasi,
\newblock \emph{Hierarchical hydrodynamics in long-range multipole-conserving systems},
\newblock Phys. Rev. B \textbf{108}, 195106 (2023),
\newblock \doi{10.1103/physrevb.108.195106}.

\bibitem{Morningstar_2023_longrange}
A.~Morningstar, N.~O’Dea and J.~Richter,
\newblock \emph{Hydrodynamics in long-range interacting systems with center-of-mass conservation},
\newblock Phys. Rev. B \textbf{108}, L020304 (2023),
\newblock \doi{10.1103/physrevb.108.l020304}.

\bibitem{Guo_2021_longrange}
Q.~Guo, C.~Cheng, H.~Li, S.~Xu, P.~Zhang, Z.~Wang, C.~Song, W.~Liu, W.~Ren, H.~Dong, R.~Mondaini and H.~Wang,
\newblock \emph{Stark many-body localization on a superconducting quantum processor},
\newblock Phys. Rev. Lett. \textbf{127}, 240502 (2021),
\newblock \doi{10.1103/physrevlett.127.240502}.

\bibitem{Scherg_2021_aidelsburger}
S.~Scherg, T.~Kohlert, P.~Sala, F.~Pollmann, B.~Hebbe~Madhusudhana, I.~Bloch and M.~Aidelsburger,
\newblock \emph{Observing non-ergodicity due to kinetic constraints in tilted {F}ermi-{H}ubbard chains},
\newblock Nat. Commun. \textbf{12}, 4490 (2021),
\newblock \doi{10.1038/s41467-021-24726-0}.

\bibitem{Kohlert_2023_aidelsburger}
T.~Kohlert, S.~Scherg, P.~Sala, F.~Pollmann, B.~Hebbe~Madhusudhana, I.~Bloch and M.~Aidelsburger,
\newblock \emph{Exploring the regime of fragmentation in strongly tilted {F}ermi-{H}ubbard chains},
\newblock Phys. Rev. Lett. \textbf{130}, 010201 (2023),
\newblock \doi{10.1103/physrevlett.130.010201}.

\bibitem{Sanchez_2020_bakr}
E.~Guardado-Sanchez, A.~Morningstar, B.~M. Spar, P.~T. Brown, D.~A. Huse and W.~S. Bakr,
\newblock \emph{Subdiffusion and heat transport in a tilted two-dimensional {F}ermi-{H}ubbard system},
\newblock Phys. Rev. X \textbf{10}, 011042 (2020),
\newblock \doi{10.1103/physrevx.10.011042}.

\bibitem{Adler_2024_2d}
D.~Adler, D.~Wei, M.~Will, K.~Srakaew, S.~Agrawal, P.~Weckesser, R.~Moessner, F.~Pollmann, I.~Bloch and J.~Zeiher,
\newblock \emph{Observation of {H}ilbert space fragmentation and fractonic excitations in {2D}},
\newblock Nature \textbf{636}, 80 (2024),
\newblock \doi{10.1038/s41586-024-08188-0}.

\bibitem{Li_2023_open}
Y.~Li, P.~Sala and F.~Pollmann,
\newblock \emph{Hilbert space fragmentation in open quantum systems},
\newblock Phys. Rev. Res. \textbf{5}, 043239 (2023),
\newblock \doi{10.1103/physrevresearch.5.043239}.

\bibitem{Ma_2019_schuster}
R.~Ma, B.~Saxberg, C.~Owens, N.~Leung, Y.~Lu, J.~Simon and D.~I. Schuster,
\newblock \emph{A dissipatively stabilized {M}ott insulator of photons},
\newblock Nature \textbf{566}, 51 (2019),
\newblock \doi{10.1038/s41586-019-0897-9}.

\bibitem{Li_2025_multiple}
L.~Li, T.~Liu, X.-Y. Guo, H.~Zhang, S.~Zhao, Z.-A. Wang, Z.~Xiang, X.~Song, Y.-X. Zhang, K.~Xu, H.~Fan and D.~Zheng,
\newblock \emph{Observation of multiple steady states with engineered dissipation},
\newblock npj Quantum Inf. \textbf{11}, 2 (2025),
\newblock \doi{10.1038/s41534-025-00958-6}.

\bibitem{Dai_2017_ringexchange}
H.-N. Dai, B.~Yang, A.~Reingruber, H.~Sun, X.-F. Xu, Y.-A. Chen, Z.-S. Yuan and J.-W. Pan,
\newblock \emph{Four-body ring-exchange interactions and anyonic statistics within a minimal toric-code {H}amiltonian},
\newblock Nat. Phys. \textbf{13}, 1195 (2017),
\newblock \doi{10.1038/nphys4243}.

\bibitem{vanNieuwenburg_2019_pnas}
E.~van Nieuwenburg, Y.~Baum and G.~Refael,
\newblock \emph{From {B}loch oscillations to many-body localization in clean interacting systems},
\newblock Proc. Natl. Acad. Sci. USA \textbf{116}, 9269 (2019),
\newblock \doi{10.1073/pnas.1819316116}.

\bibitem{Taylor_2020_moessner}
S.~R. Taylor, M.~Schulz, F.~Pollmann and R.~Moessner,
\newblock \emph{Experimental probes of {S}tark many-body localization},
\newblock Phys. Rev. B \textbf{102}, 054206 (2020),
\newblock \doi{10.1103/physrevb.102.054206}.

\bibitem{Boesl_2024_bosehubbard}
J.~Boesl, P.~Zechmann, J.~Feldmeier and M.~Knap,
\newblock \emph{Deconfinement dynamics of fractons in tilted {B}ose-{H}ubbard chains},
\newblock Phys. Rev. Lett. \textbf{132}, 143401 (2024),
\newblock \doi{10.1103/physrevlett.132.143401}.

\bibitem{Moudgalya_2020_torus}
S.~Moudgalya, B.~A. Bernevig and N.~Regnault,
\newblock \emph{Quantum many-body scars in a {L}andau level on a thin torus},
\newblock Phys. Rev. B \textbf{102}, 195150 (2020),
\newblock \doi{10.1103/physrevb.102.195150}.

\bibitem{Zerba_2024_fqh}
C.~Zerba, A.~Seidel, F.~Pollmann and M.~Knap,
\newblock \emph{Emergent fracton hydrodynamics in the fractional quantum {H}all regime of ultracold atoms},
\newblock \eprint{https://arxiv.org/abs/2410.07326}.

\bibitem{Landi_2022_boundarydrive}
G.~T. Landi, D.~Poletti and G.~Schaller,
\newblock \emph{Nonequilibrium boundary-driven quantum systems: Models, methods, and properties},
\newblock Rev. Mod. Phys. \textbf{94}, 045006 (2022),
\newblock \doi{10.1103/revmodphys.94.045006}.

\bibitem{Lidar_1998_DFS}
D.~A. Lidar, I.~L. Chuang and K.~B. Whaley,
\newblock \emph{Decoherence-free subspaces for quantum computation},
\newblock Phys. Rev. Lett. \textbf{81}, 2594 (1998),
\newblock \doi{10.1103/physrevlett.81.2594}.

\bibitem{Book_Lidar_Brun}
D.~A. Lidar and T.~A. Brun,
\newblock \emph{Quantum Error Correction},
\newblock Cambridge University Press, Cambridge, UK,
\newblock ISBN 9781139034807,
\newblock \doi{10.1017/cbo9781139034807} (2013).

\bibitem{Buca_Prosen_sym}
B.~Buča and T.~Prosen,
\newblock \emph{A note on symmetry reductions of the {L}indblad equation: transport in constrained open spin chains},
\newblock New J. Phys. \textbf{14}, 073007 (2012),
\newblock \doi{10.1088/1367-2630/14/7/073007}.

\bibitem{Baumgartner_2008_sym}
B.~Baumgartner and H.~Narnhofer,
\newblock \emph{Analysis of quantum semigroups with {GKS}–{L}indblad generators: {II. G}eneral},
\newblock J. Phys. A \textbf{41}, 395303 (2008),
\newblock \doi{10.1088/1751-8113/41/39/395303}.

\bibitem{Albert_Jiang_sym}
V.~V. Albert and L.~Jiang,
\newblock \emph{Symmetries and conserved quantities in {L}indblad master equations},
\newblock Phys. Rev. A \textbf{89}, 022118 (2014),
\newblock \doi{10.1103/physreva.89.022118}.

\bibitem{Knill_2000_NS}
E.~Knill, R.~Laflamme and L.~Viola,
\newblock \emph{Theory of quantum error correction for general noise},
\newblock Phys. Rev. Lett. \textbf{84}, 2525 (2000),
\newblock \doi{10.1103/physrevlett.84.2525}.

\bibitem{Evans_1977}
D.~E. Evans,
\newblock \emph{Irreducible quantum dynamical semigroups},
\newblock Commun. Math. Phys. \textbf{54}, 293 (1977),
\newblock \doi{10.1007/bf01614091}.

\bibitem{Katz_2023_Nbodyprogrammable}
O.~Katz, M.~Cetina and C.~Monroe,
\newblock \emph{Programmable {$N$}-body interactions with trapped ions},
\newblock PRX Quantum \textbf{4}, 030311 (2023),
\newblock \doi{10.1103/prxquantum.4.030311}.

\bibitem{Katz_2023_exp}
O.~Katz, L.~Feng, A.~Risinger, C.~Monroe and M.~Cetina,
\newblock \emph{Demonstration of three- and four-body interactions between trapped-ion spins},
\newblock Nat. Phys. \textbf{19}, 1452 (2023),
\newblock \doi{10.1038/s41567-023-02102-7}.

\bibitem{Lamata_2018_digital_sc}
L.~Lamata, A.~Parra-Rodriguez, M.~Sanz and E.~Solano,
\newblock \emph{Digital-analog quantum simulations with superconducting circuits},
\newblock Adv. Phys. X \textbf{3}, 1457981 (2018),
\newblock \doi{10.1080/23746149.2018.1457981}.

\bibitem{Fauseweh_2024_digital}
B.~Fauseweh,
\newblock \emph{Quantum many-body simulations on digital quantum computers: State-of-the-art and future challenges},
\newblock Nat. Commun. \textbf{15}, 2123 (2024),
\newblock \doi{10.1038/s41467-024-46402-9}.

\bibitem{Iaconis_2019_automaton}
J.~Iaconis, S.~Vijay and R.~Nandkishore,
\newblock \emph{Anomalous subdiffusion from subsystem symmetries},
\newblock Phys. Rev. B \textbf{100}, 214301 (2019),
\newblock \doi{10.1103/physrevb.100.214301}.

\bibitem{Pai_2019_circuits}
S.~Pai, M.~Pretko and R.~M. Nandkishore,
\newblock \emph{Localization in fractonic random circuits},
\newblock Phys. Rev. X \textbf{9}, 021003 (2019),
\newblock \doi{10.1103/physrevx.9.021003}.

\bibitem{GKS_1976}
V.~Gorini, A.~Kossakowski and E.~C.~G. Sudarshan,
\newblock \emph{Completely positive dynamical semigroups of {$N$}-level systems},
\newblock J. Math. Phys. \textbf{17}, 821 (1976),
\newblock \doi{10.1063/1.522979}.

\bibitem{Lindblad_1976}
G.~Lindblad,
\newblock \emph{On the generators of quantum dynamical semigroups},
\newblock Commun. Math. Phys. \textbf{48}, 119 (1976),
\newblock \doi{10.1007/bf01608499}.

\bibitem{Breuer_2007_book}
H.-P. Breuer and F.~Petruccione,
\newblock \emph{The Theory of Open Quantum Systems},
\newblock Oxford University Press, Oxford, UK,
\newblock ISBN 9780191706349,
\newblock \doi{10.1093/acprof:oso/9780199213900.001.0001} (2007).

\bibitem{Dhahri_2008_lindblad}
A.~Dhahri,
\newblock \emph{A {L}indblad model for a spin chain coupled to heat baths},
\newblock J. Phys. A \textbf{41}, 275305 (2008),
\newblock \doi{10.1088/1751-8113/41/27/275305}.

\bibitem{Landi_2014_lindblad}
G.~T. Landi, E.~Novais, M.~J. de~Oliveira and D.~Karevski,
\newblock \emph{Flux rectification in the quantum {$XXZ$} chain},
\newblock Phys. Rev. E \textbf{90}, 042142 (2014),
\newblock \doi{10.1103/physreve.90.042142}.

\bibitem{Schwager_2013_lindblad}
H.~Schwager, J.~I. Cirac and G.~Giedke,
\newblock \emph{Dissipative spin chains: {I}mplementation with cold atoms and steady-state properties},
\newblock Phys. Rev. A \textbf{87}, 022110 (2013),
\newblock \doi{10.1103/physreva.87.022110}.

\bibitem{Daley_2014_lindblad}
A.~J. Daley,
\newblock \emph{Quantum trajectories and open many-body quantum systems},
\newblock Adv. Phys. \textbf{63}, 77 (2014),
\newblock \doi{10.1080/00018732.2014.933502}.

\bibitem{Suleymanzade_2020_tunablecavity}
A.~Suleymanzade, A.~Anferov, M.~Stone, R.~K. Naik, A.~Oriani, J.~Simon and D.~Schuster,
\newblock \emph{A tunable high-{Q} millimeter wave cavity for hybrid circuit and cavity {QED} experiments},
\newblock Appl. Phys. Lett. \textbf{116}, 104001 (2020),
\newblock \doi{10.1063/1.5137900}.

\bibitem{Chenu_2017_noise}
A.~Chenu, M.~Beau, J.~Cao and A.~del Campo,
\newblock \emph{Quantum simulation of generic many-body open system dynamics using classical noise},
\newblock Phys. Rev. Lett. \textbf{118}, 140403 (2017),
\newblock \doi{10.1103/physrevlett.118.140403}.

\bibitem{Andrews_1984_paritions}
G.~E. Andrews,
\newblock \emph{The Theory of Partitions},
\newblock Cambridge University Press, Cambridge, UK,
\newblock ISBN 9780511608650,
\newblock \doi{10.1017/cbo9780511608650} (1984).

\bibitem{Moudgalya_2021_spectrum}
S.~Moudgalya, A.~Prem, D.~A. Huse and A.~Chan,
\newblock \emph{Spectral statistics in constrained many-body quantum chaotic systems},
\newblock Phys. Rev. Res. \textbf{3}, 023176 (2021),
\newblock \doi{10.1103/physrevresearch.3.023176}.

\bibitem{Herviou_2021_mbl}
L.~Herviou, J.~H. Bardarson and N.~Regnault,
\newblock \emph{Many-body localization in a fragmented {H}ilbert space},
\newblock Phys. Rev. B \textbf{103}, 134207 (2021),
\newblock \doi{10.1103/physrevb.103.134207}.

\bibitem{Pozderac_2023_ergidicity}
C.~Pozderac, S.~Speck, X.~Feng, D.~A. Huse and B.~Skinner,
\newblock \emph{Exact solution for the filling-induced thermalization transition in a one-dimensional fracton system},
\newblock Phys. Rev. B \textbf{107}, 045137 (2023),
\newblock \doi{10.1103/physrevb.107.045137}.

\bibitem{Classen_2024_thermalization}
J.~Classen-Howes, R.~Senese and A.~Prakash,
\newblock \emph{Universal freezing transitions of dipole-conserving chains},
\newblock \eprint{https://arxiv.org/abs/2408.10321}.

\bibitem{Huber_2020_PT}
J.~Huber, P.~Kirton, S.~Rotter and P.~Rabl,
\newblock \emph{Emergence of {$\mathcal{PT}$}-symmetry breaking in open quantum systems},
\newblock SciPost Phys. \textbf{9}, 052 (2020),
\newblock \doi{10.21468/scipostphys.9.4.052}.

\bibitem{Prosen_2012_PT}
T.~Prosen,
\newblock \emph{{$\mathbb{PT}$}-symmetric quantum {L}iouvillean dynamics},
\newblock Phys. Rev. Lett. \textbf{109}, 090404 (2012),
\newblock \doi{10.1103/physrevlett.109.090404}.

\bibitem{ElGanainy_2018_PT}
R.~El-Ganainy, K.~G. Makris, M.~Khajavikhan, Z.~H. Musslimani, S.~Rotter and D.~N. Christodoulides,
\newblock \emph{Non-{H}ermitian physics and {PT} symmetry},
\newblock Nat. Phys. \textbf{14}, 11 (2018),
\newblock \doi{10.1038/nphys4323}.

\bibitem{Agarwal_1973_detbal}
G.~S. Agarwal,
\newblock \emph{Open quantum {M}arkovian systems and the microreversibility},
\newblock Z. Phys. A \textbf{258}, 409 (1973),
\newblock \doi{10.1007/bf01391504}.

\bibitem{Znidaric_2015_gap}
M.~\u{Z}nidari\u{c},
\newblock \emph{Relaxation times of dissipative many-body quantum systems},
\newblock Phys. Rev. E \textbf{92}, 042143 (2015),
\newblock \doi{10.1103/physreve.92.042143}.

\bibitem{Can_2019_gap}
T.~Can, V.~Oganesyan, D.~Orgad and S.~Gopalakrishnan,
\newblock \emph{Spectral gaps and midgap states in random quantum master equations},
\newblock Phys. Rev. Lett. \textbf{123}, 234103 (2019),
\newblock \doi{10.1103/physrevlett.123.234103}.

\bibitem{Dutta_2021_interior}
S.~Dutta and N.~R. Cooper,
\newblock \emph{Out-of-equilibrium steady states of a locally driven lossy qubit array},
\newblock Phys. Rev. Res. \textbf{3}, L012016 (2021),
\newblock \doi{10.1103/physrevresearch.3.l012016}.

\bibitem{Ritort_2003_KCM}
F.~Ritort and P.~Sollich,
\newblock \emph{Glassy dynamics of kinetically constrained models},
\newblock Adv. Phys. \textbf{52}, 219 (2003),
\newblock \doi{10.1080/0001873031000093582}.

\bibitem{Cancrini_2007_KCSM}
N.~Cancrini, F.~Martinelli, C.~Roberto and C.~Toninelli,
\newblock \emph{Kinetically constrained spin models},
\newblock Probab. Theory Relat. Fields \textbf{140}, 459 (2007),
\newblock \doi{10.1007/s00440-007-0072-3}.

\bibitem{Garrahan_2011_KCM}
J.~P. Garrahan, P.~Sollich and C.~Toninelli,
\newblock \emph{Kinetically constrained models},
\newblock In \emph{Dynamical Heterogeneities in Glasses, Colloids, and Granular Media}, p. 341–369. Oxford University Press, Oxford, UK,
\newblock \doi{10.1093/acprof:oso/9780199691470.003.0010} (2011).

\bibitem{Gopalakrishnan_2018_automata}
S.~Gopalakrishnan and B.~Zakirov,
\newblock \emph{Facilitated quantum cellular automata as simple models with non-thermal eigenstates and dynamics},
\newblock Quantum Sci. Technol. \textbf{3}, 044004 (2018),
\newblock \doi{10.1088/2058-9565/aad759}.

\bibitem{Mukherjee_2024_hyperuniformity}
A.~Mukherjee, D.~Tapader, A.~Hazra and P.~Pradhan,
\newblock \emph{Anomalous relaxation and hyperuniform fluctuations in center-of-mass conserving systems with broken time-reversal symmetry},
\newblock Phys. Rev. E \textbf{110}, 024119 (2024),
\newblock \doi{10.1103/physreve.110.024119}.

\bibitem{Hazra_2025_hyperuniformity}
A.~Hazra, A.~Mukherjee and P.~Pradhan,
\newblock \emph{Hyperuniformity in mass transport processes with center-of-mass conservation: some exact results},
\newblock J. Stat. Mech. \textbf{2025}, 023201 (2025),
\newblock \doi{10.1088/1742-5468/ada88c}.

\bibitem{Dutta_2020_center}
S.~Dutta and N.~R. Cooper,
\newblock \emph{Long-range coherence and multiple steady states in a lossy qubit array},
\newblock Phys. Rev. Lett. \textbf{125}, 240404 (2020),
\newblock \doi{10.1103/physrevlett.125.240404}.

\bibitem{Book_NIST}
F.~W.~J. Olver, D.~W. Lozier, R.~F. Boisvert and C.~W. Clark, eds.,
\newblock \emph{NIST Handbook of Mathematical Functions},
\newblock Cambridge University Press, New York (2010).

\end{thebibliography}


\end{document}